\documentclass{article}

\begin{document}

\title{\bf Geometric phase from Dielectric matrix}
\author{Banerjee Dipti \\ Department of Physics \\
Rishi Bankim Chandra College \\
Naihati,24-Parganas(N)\\
Pin-743 165, West Bengal\\
INDIA}
\maketitle

\abstract
The dielectric property $(2\times2)$ of the anisotropic optical
medium is found out considering the polarized photon as two
component spinor of spherical harmonics.The Geometric Phase of
single polarized photon has been evaluated in two ways. The phase
two-form of the dielectric matrix through a twist and the
Pancharatnam phase(GP)through the change of angular momentum of
the incident polarized photon over a closed triangular path on the
extended Poincare sphere.The helicity in connection with the spin
angular momentum of the chiral photon plays the key role in
developing these phase holonomies.

\vspace{2cm}

email: deepbancu@hotmail.com
\newpage

\section{\bf Introduction}

The idea of geometric phase was first pointed out theoretically in
1941's by Vladimirski[1]in connection with curved nonplaner light
beam for its rotation of plane of polarization in an inhomogenous
medium.In 1956 Pancharatnam showed similar phase[2] in his
strikingly original work on interference of polarized light by the
cyclic change of state over a closed path on the Poincare sphere.
The similarity of the above work with Berry [3] was pointed out by
Ramseshan et.al[4].This leads Berry to study the {\it Geometric
Phase}(GP) of linearly polarized light from quantum mechanical
point of view[5].Twisting the anisotropic medium over a closed
path, Berry also[6] studied the phase two-form(GP)in connection
with the dielectric tensor and bierfringence of the medium.The
geometric phase in the context of polarization optics have been
studied from the view-point of group theoretical aspects by Simon
and Mukunda [7] and Bhandari[8].

The physical mechanism for GP in Optics has been elucidated in
terms of angular momentum holonomy  by Tiwari [9].Recently Galvez
et.al.[10] has given a experimental measurement of the geometric
phase acquired by the Optical beams bearing orbital angular
momentum. They observe the GP in mode space as suggested by van
Enk [11].Now a days the angular momentum of Photon takes a new way
to the physical approach. More recently Leach et.al [12] proposed
a interferometric methods for measuring the spin and orbital
angular momentum of single photon.

All these recent findings indicate that our previous study on
Geometric phase of a polarized photon (passing through the
polarization matrix-M and a rotator) in connection with {\it
helicity}[13] was an obvious new representation of GP.We have
extended this idea evaluating GP through differential matrix N
from-M [14] In the light of Jones work[15],we in this paper will
study the GP in connection with the dielectric properties of the
optical media. The dielectric property of M is determined from N
matrix to evaluate first the phase two-form.Then the change of
polarization of the incident light is made over a closed triangle
by a rotator such a way that the variation of angular momentum is
visualized through the{\it Pancharatnam Phase} in the relativistic
framework.

\section{\bf The spinorial representation of polarized quanta.}

 Propagation through distorting obstacle significantly
influence the amplitude, phase and polarization state of a light
beam.This has consequences for the behavior of the orbital angular
momentum (OAM) of Optical beam that has been measured
experimentally [10]. Berry pointed out[16] that Photons have no
magnetic moment and so cannot be turned with a magnetic field. But
they have the property of the helicity, along their propagation
direction $e_k$. They may have states with two {\it helicities}
$(\sigma.e_{k})=\pm1$, but not zero.The eigenstates of helicity
with eigenvalues $+1$ and $-1$ are referred to, respectively, as
the {\it right-handed} state (spin parallel to motion) and the
{\it left-handed} state (spin opposite to motion).

This indicates that the photon in the polarized beam fixes its
{\it helicity} whose direction changes with the change
  of polarization. In an anisotropic space a particle
   having a fixed {\it helicity} can be viewed as if a {\it direction vector}
   is attached at the space-time point[17].From the relativistic point of view, if
    $x_\mu$ is the mean position of the particle and $y_\mu$ indicates the {\it direction vector}
      then we can consider the resultant coordinate in the complexified space
 as $z_\mu =x_\mu +iy_\mu$. This extended structure indicates the acquirance
of mass and masslessness condition is achieved when we have
$|y_\mu|^2=0$. It can be shown that the two opposite orientations
of the {\it direction vector} represent two {\it internal
helilities} corresponding to fermion and antifermion. In view of
this, we can formulate the {\it internal helicity} in terms of two
component spinorial variables $\theta(\bar{\theta})$.

Indeed, in the complexified space-time we can write the chiral
coordinate as [17]
\begin{equation}
z^\mu = x^\mu +iy^\mu =
     = x^\mu +(i/2){\lambda_\alpha}^\mu  \theta_\alpha
\end{equation}
If we now replace the chiral coordinates by their matrix
representations
\begin{equation}
z^{AA^\prime}=
x^{AA^\prime}+(i/2){\lambda_\alpha}^{AA^\prime}\theta^\alpha
\end{equation}
where
\begin{equation}
x^{AA^\prime} = \left(
\begin{array}{rr}
x^0-x^1 & x^2+ix^3 \\ x^2-ix^3 & x^0+x^1
\end{array} \right)
\end{equation}
with
$${\lambda_\alpha}^{AA^\prime} \in SL(2,c)$$ This involves the {\it helicity} operator
\begin{equation}
S_{hel} =- {\lambda_\alpha}^{AA^\prime} {\theta}^\alpha {\bar
\Pi_A}\Pi_A^\prime
\end{equation}

which we identify as the {\it internal helicity} and  corresponds
to the fermion number when the two opposite orientations of {\it
internal helicities} represent particle and antiparticle. It may
be noticed that  we have taken the matrix representation of
$p_\mu$ (conjugate to $x_\mu$ in the complex coordinate
$z_\mu=x_\mu+iy_\mu$) as $p^{AA^\prime} =
{\bar\pi^A}\pi^{A^\prime}$ when ${p_\mu}^2 = 0$.  So the particle
will have its mass due to the nonvanishing character of the
quantity ${y_\mu}^2$. It is observed that the complex conjugate of
the chiral coordinate (1) will give rise to a massive particle
with opposite
  {\it internal helicity} corresponding to an antifermion.
   In the null plane where ${y_\mu}^2=0$, we can write the
    chiral coordinate for
    massless spinor as follows
\begin{equation}
z^{AA^\prime}=x^{AA^\prime}+\frac{i}{2}{\bar \theta}^A
\theta^{A^\prime}
\end{equation}

where the coordinate $y^\mu$ is replaced by
$y^{AA^\prime}=(1/2){\bar \theta }^A \theta ^{A^\prime}$. In this
case the {\it helicity} operator is given by
\begin{equation}
S=-{\bar \theta}^A {\theta}^{A^\prime} {\bar \pi}_A
\pi_{A^\prime}=-{\bar \varepsilon}\varepsilon
\end{equation}
where $\varepsilon=i\theta^{A^\prime} \pi_ A$,${\bar
\varepsilon}=-i{\bar \theta} ^A\pi_{A^\prime}$. The corresponding
twistor equation describes a massless spinor field.

In case of massive spinor, we can define a plane $D^-$, where for
coordinate $z_\mu=x_\mu +iy_\mu $, $y_\mu $ belongs to the
interior of forward lightcone $ (y>>0)$ and as such represents the
upper half-plane with the condition det$y^{AA^\prime}>0$ and (1/2)
Tr$y^{AA^\prime}>0$. The lower half plane $D^+$ is given by the
set of all coordinates $z_\mu$ with $y_\mu$ in the interior of the
backward lightcone $(y<<0)$. The map  $z \rightarrow z^*$ sends
the upper half-plane to the lower half-plane. The space M of null
planes $(det y^{AA^\prime}=0)$ is the Shilov boundary so that a
function holomorphic in $D^{-}(D^{+})$ is determined by its
boundary values. Thus,if we consider that any function
$\phi(z)=\phi(x)+i\phi(\xi)$ is holomorphic in the whole domain
the {\it helicity} $+\frac{1}{2}(-\frac{1}{2})$ in the null plane
may be taken to be the limiting value of the {\it internal
helicity} in the upper(lower) half-plane. Thus massless spinor
exists in this plane. For a massive particle {\it helicity} is
incorporated in the internal space and identified as {\it internal
helicity} which introduces solitonic feature for the spinor and
gives rise to a massive fermion. In view of this, the {\it
internal helicity} may be taken to represent the fermion number
where {\it helicity} is associated with spin for massless fermion.

It may be noted that wave-function
$\phi(z_\mu)=\phi(x_\mu)+i\phi(y_\mu)$ can be treated to describe
a particle moving in the external space-time having the coordinate
$x_\mu$ with attached{\it direction vector} $y_\mu$. Thus the wave
function should take into account the polar coordinates $r$,
$\theta, \phi$ along with the angle $\chi$ which specify the
rotational orientation around the {\it direction vector}
$y_\mu$.for an extended particle $\theta, \phi$ and $\chi$ just
represent the three Euler angles.

In a $3D$ anisotropic space, we can consider an axis symmetric
system where the anisotropy is introduced along a particular
direction. It is to be noted that in this anisotropic space, the
components of linear momentum satisfy a commutation relation of
the form
\begin{equation}
[p_i,p_j]= i\mu \epsilon_{ijk}\frac{x^k}{r^3}
\end{equation}

In such a space,the conserved angular momentum $J$ represented by
\begin{equation}
\vec{J}=\vec{r}\times \vec{p} - \mu\vec{r}
\end{equation}
It follows that $J^2=L^2-\mu^2$ instead of $L^2$ is a conserved
quantity. In general,$\mu$ which is the measure of the anisotropy
given by the eigenvalue of the operator
$i\frac{\delta}{\delta_\chi}$ and can take the values $\mu=0,\pm
1/2,\pm 1,\ldots$.The spherical harmonics incorporating the term
$\mu$ be written as [15]
\begin{equation}
{Y_l}^{m,\mu}=(1+x)^{-(m-\mu)/2}\,(1-x)^{-(m+\mu)/2}
\frac{d^{l-m}}{{dx}^{l-m}}\,[(1+x)^{l-\mu}(1-x)^{l+\mu}\,]\times
e^{im\phi}e^{-i\mu\chi}
\end{equation}
with $x=cos\theta$.

In the anisotropic space a scalar particle moving with $l=1/2$
with $l_z=+1/2$ can be treated as a spinor with {\it helicity}
$+1/2$.The specification of the $l_z$ value for the particle and
antiparticle states then depicts it as a chiral spinor. This gives
for $m=\pm 1/2,\mu=\pm 1/2,$the following spherical harmonics from
the relation (9) in terms of the components $(\theta,\phi,\chi)$
\begin{equation}
\left.
\begin{array}{lcl}
{Y_{1/2}}^{1/2,1/2} & = & \sin \frac{\theta}{2} e^{i(\phi-\chi)/2} \\
{Y_{1/2}}^{-1/2,1/2} & = & \cos \frac{\theta}{2} e^{-i(\phi+\chi)/2} \\
{Y_{1/2}}^{1/2,-1/2} & = & \cos \frac{\theta}{2} e^{i(\phi+\chi)/2} \\
{Y_{1/2}}^{-1/2,-1/2} & = & \sin \frac{\theta}{2}
e^{-i(\phi-\chi)/2}
\end{array}
\right\}
\end{equation}

These represent spherical harmonics for half-orbital angular
momentum in an anisotropic space and it is to be noted that from
these spherical harmonics we can construct the polarization matrix
$M$ of the complete optical system [13]. Here, in the next section
we shall use this $M$ and $N$ to evaluate the dielectric
properties of the optical device. Then GP of single photon will be
studied from two different approaches in the relativistic
framework.

\section{\bf The matrix formulation of the Dielectric matrix
of the optical system and the geometric phase}

The passage of plane polarized light through partial polarizer or
retardation plate having its plane of polarization parallel to
either of the principal axes, suffer no change in the state of
polarization. Similarly both type of circularly polarized light
acquire no change in their state of polarization in passing
through a rotator. This property of light with respect to the
optical element was presented in $2\times2$ matrix method by
Jones[18]. The matrix $M^n$ of the optical system of n components
having eigenvectors $\varepsilon_i$ satisfy the condition
\begin{equation}
M^n \varepsilon_i=d_i\varepsilon_i
\end{equation}
where $d_i$ is the constant known as eigenvalue corresponding to
the eigenvectors $\varepsilon_i$. For one component  optical
element the matrix becomes
$ M=\pmatrix{m_1&m_4\cr m_3&m_2}.$\\

Jones has pointed out that the matrix  $M$ of the optical system
is then uniquely determined by the relation
\begin{eqnarray}
M&=&TDT^{-1} \\
M&=&\Delta^{-1}\pmatrix{{d_1a_1b_2 -
d_2a_2b_1}&{-(d_1-d_2)a_1a_2}\cr
{(d_1-d_2)b_1b_2}&{d_2a_1b_2-d_1a_2b_1}}
\end{eqnarray}
from the orthonormal eigenvectors $\varepsilon_1$, $\varepsilon_2$
and eigen values $D$
\begin{eqnarray}
\varepsilon_1&=&a_1\choose b_1 \\ \varepsilon_2&=&{-b_1}^*\choose
{a_1}^*
\end{eqnarray}
$$\Delta=a_1b_2 - a_2b_1$$  $$D=\pmatrix{d_1&0\cr 0&d_2}$$
When the optical system is a homogeneous material, one can
describe the optical properties (dielectric and gyration) of the
device by the differential matrix $N$, such that it is related
with the matrix operator $M$ of the complete element as follows
\begin{equation}
N=\frac{dM}{dz}M^{-1}
\end{equation}
where $$ N=\pmatrix{n_1&n_4 \cr n_3&n_2}$$ and $$M_z =expN_z$$ The
propagation through an infinitesimal distance however is described
by the Matrix $N$,the vibration at the plane $z+dz$ of the medium,
for a given direction and for a given wave-length, may be defined
as the linear vector function of the vibration at the plane $z$ by
the relation
\begin{equation}
\frac{dD}{dz}=ND
\end{equation}
The propagation of polarized light [18] through the anisotropic
homogenous medium can be represented by the displacement vector
that satisfy the following relation
\begin{equation}
\frac{\partial^2 D}{\partial t^2}=c^2 [n]^{-2}\frac{\partial^2
D}{\partial z^2}
\end{equation}
where $[n]$ is the refractive index matrix of the medium. This
resembles the usual form of wave equation where the velocity is
replaced by the refractive index matrix $[n]$. For homogenous
anisotropic medium [16],we have$[n]^2=[\varepsilon]$
  where $[\varepsilon]$ is the two by two  matrix representing the
   dielectric tensor of the medium  connecting $\vec{D}$.
    It satisfies the customary wave equations of homogenously polarized light
representing the states $\vec{D}$ by the following relations
\begin{equation}
[\varepsilon]\vec{D}= n^2\vec{D}
\end{equation}
where eigenvalues are the refractive indices of the medium. Jones
[19] had shown that the two component dielectric tensor can be
written in the terms of  differential matrix $N$ of the material
by
\begin{equation}
[\varepsilon]= -\left(N^2 + \frac{dN}{dz}\right)
\end{equation}
We will study here the geometric phase (GP) from the dielectric
property of the polarization matrix $M$[13].This GP is concerned
with the rotation of the polarized light over a closed path by
twisting once the optical medium about the direction of
propagation.
\begin{equation}
\gamma = arg<\psi_{initial}|\psi_{final}>
\end{equation}
In case the polarization of the incident light changes over a
closed path on the Poincare sphere, then the GP is identified as
Pancharatnam phase.
 Photons have no magnetic moment and so cannot be
turned with a magnetic field. But they have the property of the
helicity, along their propagation direction $e_k$ They may have
states with two {\it helicities} $(\sigma.e_{k})=\pm1$, but not
zero. The geometric phase the two helicities of polarized photon
would be [3]
\begin{equation}
\gamma_{\pm1}(c)=\pm \Omega(c)
\end{equation}
where $\gamma$ is the solid angle  swept out by $e_k$ on the
sphere.These two helicities for polarization states originates
from the spin angular momentum which has created significant
interest in recent years. Hence our emphasis on the fixed{\it
helicity} of polarized photon is fruitful in evaluating GP.

With the above idea the polarization matrix $M$ of the optical
element have been calculated recently [13].
\begin{equation}
\begin{array}{lcl}
M=TDT^{-1}=1/2\pmatrix{-\cos\theta & \sin\theta e^{-i\chi}\cr
\sin\theta e^{i\chi} & \cos\theta \cr}
\end{array}
\end{equation}
represented by each point on the Poincare sphere parameterized by
the angle $\theta$ and $\chi$. Here $T$ is the matrix formed by
the orthonormal eigenvecters of spherical harmonics and $D$
denotes the eigenvalue matrix that reflects here the helicity
(+1/2,-1/2) of the polarized photon.Here we have represented (in
sec. 2) relativistically the polarized photon by two component
spinor of spherical harmonics where the effect of {\it helicity}
is visualized by the parameter $\chi$. The behavior of chiral
photon with a fixed {\it helicity} $\pm 1$ in the polarized light
is similar to massless fermion having {\it helicity} +1/2 or -1/2.

At a particular position of $z$, the $N$ matrix is related with
$M$ as follows
\begin{equation}
N=(\frac{dM}{d\theta})(\frac{d\theta}{dz})M^{-1}
\end{equation}
where $z=cos\theta$.
 Substituting the matrix value of $M$ in the
above equation, the $N$ matrix is obtained
\begin{equation}
N=1/{2\sin\theta} \pmatrix{0 & e^{-i\chi} \cr -e^{i\chi} & 0}
\end{equation}

Finally  from the obtained differential matrix $N$[14], the two
component dielectric matrix  $[\varepsilon]$ has been calculated
\begin{equation}
[\epsilon]=1/4 \pmatrix{{\rm cosec}^2\theta & - 2 {\rm cot}\theta
e^{-i\chi} \cr 2 {\rm cot}\theta e^{i\chi} & {\rm cosec}^2\theta}
\end{equation}
This dielectric matrix resides on the same Poincare sphere of
polarization matrix $M$ parameterized by the two variables
$\theta$ and $\chi$. This latter measures the angle between the
direction of measurement and the spin angular momentum of the
polarized photon where the amount of anisotropy $\mu$ is
visualized through the relation $i\frac{\partial}{\partial \chi}$.
Our discussions in previous section suggests that variation of the
angle $\chi$ is associated with the change of angular momentum
through the factor $\mu$. We thus add here following the views of
our previous work [13] that the change of polarization of light is
in connection with the change of angle $\chi$. And this is
possible when light passing through the optical devices suffer
change of polarization due to transfer of angular momentum. This
view is similar to that of Tiwari [9].

Berry [6] showed that the classical optics of slowly-varying
dielectric media with both bierfringence and gyrotropy can
generate geometrical phase shifts. The phase 2-form for such $H$
matrices may be denoted by $V_{\pm}(x)$ where $\pm$ refers to the
eigenstate with the higher/lower eigenvalue. There is a useful
general formula for this matrix in the form
\begin{equation}
H(x)=\pmatrix{A_0+A_z & A_x- i A_y \cr A_x+ i A_Y & A_0-A_z}
\end{equation}
whose cyclic change of parameter corresponds to a circuit on the
Poincare sphere which is accompanied by the phase 2-form
$V_{\pm}(x)$ known by
\begin{equation}
V_{\pm}(x)=\frac{A_x dA_y \wedge dA_z + A_y dA_z \wedge dA_x + A_z
dA_x \wedge dA_y}{2({A^2}_x+{A^2}_y +{A^2}_z)^{3/2}}
\end{equation}
In the light of Berry's work,the dielectric matrix $[\varepsilon]$
is analogous to $H$, where $A_0=0$, the $A_x=2{\rm cot}\theta
\cos\chi$, $A_y=2{\rm cot}\theta \sin\chi$ and $A_z={\rm
cosec}^2\theta$, only differs from the sign of the determinant.
Here the anisotropy in the dielectric matrix $[\varepsilon]$
 combine the unaxial birefringence or gyratropy
being present in the phase 2-form $V(\theta,\chi)$ which will give
rise a nonzero $\gamma$ or GP.
\begin{equation}
V_{\pm}(\theta,\chi)=\pm \frac{\sin2\theta \cos2\theta}{(1+
{\sin^2 2\theta}) ^{3/2}} d\theta\wedge d\chi
\end{equation}
Now the plane of polarization is turned so as to sweep out a cone,
by keeping $\theta$ constant at a particular $z$ values and
increasing $\chi$ by $2\pi$. This rotation of $\chi$ over a closed
path implies the rotation of helicity of the polarized photon
whose net chiral change develops our required relativistic
geometric phase.
\begin{equation}
\gamma(\theta)=\pm 4\pi\left(1- \frac{1}{\sqrt{1 + \sin^2 2
\theta}}\right)
\end{equation}
Thus it is possible to find out a nonzero GP
($\gamma(\theta)$)depending on the angle $\theta$ which is the
phase two-form obtained from our dielectric matrix.At $\theta=0$
and $\pi/2$ the value of $\gamma(\theta)$ becomes zero, whereas it
attains the maximum value $4\pi(1-1/\sqrt{2})$ for
$\theta=\pi/4$.Further we like to add that this GP is not
visualizing the spin angular momentum transfer through the
variable $\chi$ parameterizing helicity. Next we will proceed to
evaluate the Pancharatnam Phase through variation of {\it
helicity}.

If the dielectric variation is slow enough, the beam remains in a
polarization state determined by the local uniform medium. The
displacement vector is one of the two eigenvectors of the
$(2\times2)$ hermitian sub-matrix of the dielectric tensor.In view
of this, the dielectric matrix satisfying the eigenvalue equation
$$[\epsilon] \vec{D}=n^2 \vec{D}$$
having eigenvectors
\begin{equation}
\vec{D}={\rm cot} \theta {\pm i \choose e^{i\chi}}
\end{equation}
with eigenvalues $1/4({\rm cosec}^2\theta \pm 2i \rm cot\theta)$
identify the complex refractive index $n$ of the anisotropic
homogenous medium.

The effect of rotator is introduced in the optical system such a
way that it rotates the plane of polarization by an angle $\phi/2$
about the z axis. Then the resultant optical system becomes
\begin{equation}
M'= S(\phi/2) [\epsilon ] S(-\phi/2)
\end{equation}
in the following matrix form
\begin{equation}
M'=\pmatrix{cos\phi/2 & -sin\phi/2 \cr sin\phi/2 & cos\phi/2} 1/4
\pmatrix{{\rm cosec}^2\theta & - 2 {\rm cot}\theta e^{-i\chi} \cr
2 {\rm cot}\theta e^{i\chi} & {\rm cosec}^2\theta}
\pmatrix{cos\phi/2 & sin\phi/2 \cr -sin\phi/2 & cos\phi/2}
\end{equation}
It seems that the geometric surface developed by the polarization
matrix $M'$ is parameterized by three variables $\theta, \phi$ and
$\chi$ (as similar as Euler angles). The usual polarization matrix
$(2\times 2)$ developed by the parameters $\theta$ and $\phi$ lies
on the Poincare sphere $S^2$. Here $M'$ resides on the extended
Poincare sphere. As the plane of polarization of the incident
polarized light is made to change over a closed path GP is
developed.Since this phase is developed due to the change of
angular momentum for variation of $\chi$ it can be termed as
angular momentum holonomy.
 We consider the rotation of the optical system represented
by the Dielectric matrix followed by the rotator over the contour
of a triangle on the extended Poincare sphere, whose vertices are
$A=(\pi/4,0),B=(\pi/4,\pi/2)$ and $C= (\pi/2,\pi/2)$.This path ABC
encloses a triangular area, as the eigen spinor of the point A
\begin{equation}
|\psi(A)>={\rm{i\choose e^{i\chi}}}
\end{equation}
reaches its initial points through $B=(\pi/4,\pi/2)$ and
 $C=(\pi/2,\pi/2)$, a development of the geometric phase is
visualized by
\begin{equation}
<\psi(A)|\psi(C)>= <\psi(A)|M'_C M'_B M'_A|\psi(A)>
\end{equation}
It can be noted that rotated dielectric matrices at the respective
points $A=(\pi/4,0),B=(\pi/4,\pi/2) and C= (\pi/2,\pi/2)$ are
\begin{equation}
M'(A)= 1/2\pmatrix{1 & -e^{-i\chi} \cr e^{i\chi}  & 1} \\
M'(B)= 1/2\pmatrix{1 & -e^{i\chi} \cr e^{-i\chi}  & 1} \\
M'(B)= 1/4\pmatrix{1 & 0 \cr 0  & 1}
\end{equation}
After a few mathematical steps we obtain the geometrical phase of
a single photon as
\begin{equation}
\gamma=4i\cos^2{\chi} - 2\cos{2\chi}+2
\end{equation}

 As the plane of polarization of the emergent light from the
 dielectric matrix varies over a closed path by a rotator,
the Geometric phase is developed. Since the GP has only dependence
on the parameter $\chi$ indicating the angular momentum variation
or the change of anisotropy through $\mu$, this phase can be
identified as the angular momentum holonomy. It does not include
any dynamical phase dependence through the parameter
$\theta$,$\phi$. Hence the dynamical phase can be automatically
removed without using `spin-echo' method[20] by passing the
emergent plane polarized light from the dielectric matrix through
a rotator. We suggest the experimental application of this work to
evaluate the angular momentum holonomy through the approach of
Pancharatnam phase.

{\bf Discussions}

I infer at the end that in the light of Jones calculus, I have
determined the dielectric properties ($2\times 2$ matrix) of the
polarization matrix M[13]. We have found these matrices on the
Poincare sphere parameterized by the variables $\theta$ and $\chi$
where the latter defines the inclination of {\it helicity} of the
polarized photon. The orthogonal eigen spinors that are
responsible for the construction of the above matrices are the two
component spinor of spherical harmonics ${Y_l}^{m,\mu}$. We have
shown the appearance of the Geometric phase in two different
ways.\\
 1. The phase two-form of the Dielectric matrix by a twist.\\
 2. The rotation of the plane of polarization by inserting
      a rotator along with the optical device. The polarized light
      is made to follow a triangular closed path on the extended
      Poincare sphere ($\theta, \phi and \chi$).

The phase two-form is obtained by turning once the
 polarization (dielectric) matrix at a fixed $\chi$ over a closed
 path on the Poincare sphere ($\theta, \chi$).
 In the latter case, the polarization of the incident light is made to change
over a closed triangle on the extended Poincare sphere
parameterized by $\theta,\phi and \chi$. The variable $\chi$ is
associated with the {\it helicity} or plane of polarization and it
changes during the course of rotation.Though the final state
coincide with the initial their remains a development of angular
momentum holonomy (GP) by the change of angular momentum through
the parameter $\chi$ of the chiral photon.The advantage of having
the GP (Pancharatnam phase) of a single photon through the
dielectric matrix and rotator is that it fully depends on {\it
helicity} of spin angular momentum through parameter $\chi$. Hence
dynamical phase can be removed without using `spin-echo' method
[20]. This is a new relativistic approach to evaluate
experimentally the angular momentum holonomy of the polarized
photon.

{\bf Acknowledgement}

I express my gratitude to all the authors in my references
specially Dr. Tiwari for his helpful correspondence and comments
and Prof.P. Bandyopadhyay for his encouragement.


\section{\bf Reference}

\begin{enumerate}

\item[1] V.V Vladimirski;Dokl.Acad.Nauk.{\bf 31},31 (1941).in
{\it Topological Phases in quantum Theory}, Ed: B. Markovski and
S.I.Vinitsky (World Scientific Publishing Co; 1989).\\
\item[2] S.Pancharatnam; Proc.Ind.Acd.Sci;{\bf A44},247,(1956).\\
\item[3] M.V.Berry, Proc.Roy.Soc.London;{\bf A392},45,(1984).\\
\item[4] S.Ramaseshan and R. Nityananda; Current Science,India,{\bf 55},1225,(1986).\\
\item[5] M.V.Berry;J.Mod.Opt:{\bf 34},1401,(1987). \\
\item[6] M.V.Berry;Proceedings of a NATO Advanced Research
Workshop on Fundamental Aspects of Quantum Theory, held in Italy, Sep. 1985;\\
\item[7] R.Simon and N.Mukunda; Phys.Lett{\bf 138A},474,(1989).\\
\item[8] R.Bhandari; Phys. Lett{\bf157},221,(1989).\\
\item[9] Tiwari.S.C; J.Mod.Opt.vol-{\bf 39},(1992),1097;J.Opt.B,{\bf 4},2002, S39;
physics/0310091.\\
\noindent[10] Galvez E.J.,Crawford R.R,Sztul H.I, Pysher M.J,and Williams R.E,
 Phys.Rev.Lett, {\bf 90},(2003),203901.\\
\item[11] S J van Enk,Optics commun,{\bf 102}.59 (1993).\\
\item[12] J Leach et al PRL {\bf 92},013601 (2004).\\
\item[13] D.Banerjee; Phys.Rev.-{\bf 56E},1129,(1997).\\
\item[14] D.Banerjee; The Study of Geometric Phase with Twisted Crystal (communicated).\\
\item[15] R.C.Jones; J.Opt.Soc.Am.{\bf 31} 488 (1941);\\
\item[16] M.V.Berry; Lectures presented at the International
School on "Anomalies, Phases, Defects...held in Ferrara, Italy,June 1989.\\
\item[17] P.Bandyopadhyay; Int.J.of Mod.Phys: {\bf A4},
4449,(1989);{\it Geometry,Topology and Quantization}(Kluwer
Academic Publisher, 1996, The Netherlands).\\
\item[18] R.C.Jones; J.Opt.Soc.Am{\bf 38},671,(1948).\\
\item[19] R.C.Jones; J.Opt.Soc.Am,{\bf 46},126,(1956).
 G.N Ramachandran and S.Ramaseshan, {\it Crystal Optics in
Encyclopedia of Physics},Ed.S.Flugge, vol,25,Part-I,
(Springer-Verlag;1961)\\
\item[20] R.A.Bertlmann, K.Dustberger,Y.Hasegawa and
B.C.Hiesmayr; quantut-ph/0309089,(2003).\\
\end{enumerate}

\end{document}